\documentclass{article}
\usepackage{spconf,amsmath,graphicx}
\usepackage{amssymb,amsmath,amsthm,enumitem}
\usepackage[pagebackref,breaklinks,colorlinks]{hyperref}
\usepackage{enumitem}
\setlist{nosep, leftmargin=14pt}

\usepackage{mwe} 
\usepackage{booktabs}
\usepackage{pifont}%
\usepackage{soul}

\newcommand{\cmark}{\ding{51}}%
\newcommand{\xmark}{\ding{55}}%

\title{Author guidelines for ISBI proceedings manuscripts}
\title{TSRNet: Simple Framework for Real-time ECG Anomaly Detection with Multimodal Time and Spectrogram Restoration Network}
\name{Nhat-Tan Bui$^{1*}$, Dinh-Hieu Hoang$^{2,3*}$, Thinh Phan$^{1}$, Minh-Triet Tran$^{2,3}$,}{Brijesh Patel$^{4}$, Donald Adjeroh$^{4}$, Ngan Le$^{1}$}
\address{
$^1$AICV Lab, University of Arkansas, Arkansas, USA\\
$^2$University of Science, Vietnam National University, Ho Chi Minh City, Vietnam \\
$^3$John von Neumann Institute, Vietnam National University, Ho Chi Minh City, Vietnam\\
$^4$West Virginia University, West Virginia, USA \\
}
\begin{document}
%
\maketitle
\def\thefootnote{*}\footnotetext{Equal contribution}\def\thefootnote{\arabic{footnote}}
\begin{abstract}
The electrocardiogram (ECG) is a valuable signal used to assess various aspects of heart health, such as heart rate and rhythm. It plays a crucial role in identifying cardiac conditions and detecting anomalies in ECG data. However, distinguishing between normal and abnormal ECG signals can be a challenging task. In this paper, we propose an approach that leverages anomaly detection to identify unhealthy conditions using solely normal ECG data for training. Furthermore, to enhance the information available and build a robust system, we suggest considering both the time series and time-frequency domain aspects of the ECG signal. As a result, we introduce a specialized network called the Multimodal \textbf{T}ime and \textbf{S}pectrogram \textbf{R}estoration Network (TSRNet) designed specifically for detecting anomalies in ECG signals. TSRNet falls into the category of restoration-based anomaly detection and draws inspiration from both the time series and spectrogram domains. By extracting representations from both domains, TSRNet effectively captures the comprehensive characteristics of the ECG signal. This approach enables the network to learn robust representations with superior discrimination abilities, allowing it to distinguish between normal and abnormal ECG patterns more effectively. Furthermore, we introduce a novel inference method, termed Peak-based Error, that specifically focuses on ECG peaks, a critical component in detecting abnormalities. The experimental result on the large-scale dataset PTB-XL has demonstrated the effectiveness of our approach in ECG anomaly detection, while also prioritizing efficiency by minimizing the number of trainable parameters. Our code is available at 
\url{https://github.com/UARK-AICV/TSRNet}.
\end{abstract}

\section{Introduction}
\vspace{-0.5em}
The electrocardiogram (ECG) is a crucial time series data that provides valuable insights into the detection and diagnosis of cardiac diseases. The morphology observed in each lead of ECG provides insight into the electrical activity occurring within specific segments of the heart. This information proves invaluable in identifying abnormal myocardial conditions \cite{davies2014starting, phan2022multimodality, le2021multi, le2023scl}. However, focusing solely on the classification of specific diseases through ECG analysis may result in the failure to detect various types of abnormal signals that were not encountered during training. Therefore, our attention shifts towards the detection of anomalies in ECG signals, a strategy with the potential to identify abnormal cardiac conditions in patients, regardless of the diversity and rarity of these conditions.

Unsupervised Deep Learning methods such as \cite{dagmm, madgan, usad, anotran, mscred, gdn, cae-m, tranad} have demonstrated significant success in the realm of time series anomaly detection. Currently, BeatGAN \cite{beatgan}, and Jiang et al. \cite{ecg-miccai}, which fall under the restoration-based (also called reconstruction-based) category, are primarily designed for ECG anomaly detection. Although these methods have attained notable advancements, they exclusively utilize the 1D ECG signals (i.e. time series domain), overlooking the potential of the 2D ECG spectrum (i.e. time-frequency domain). The ECG spectrogram, which represents meaningful time-frequency representation, has been successfully applied in ECG classification tasks \cite{sun2022automatic, phan2022multimodality, le2021multi}. Therefore, exploring the incorporation between time-series and spectrogram offers an alternative avenue for tackling the ECG anomaly detection challenge. As a result, this paper introduces Multimodal \textbf{T}ime and \textbf{S}pectrogram \textbf{R}estoration Network (\textbf{TSRNet}), an unsupervised anomaly detection method designed for ECG anomaly detection. TSRNet combines both 1D ECG time series and 2D ECG spectrogram data using attention modules. The primary
goals of our method are to extract and fuse meaningful features from both domains to robustly detect anomalies in ECG signals while maintaining simplicity and parameter efficiency. Our contributions include investigating the potential benefits of spectrograms in ECG anomaly detection, developing the TSRNet model, which incorporates a novel Peak-based Error, and evaluating its performance through multiple experiments on the PTB-XL dataset \cite{ptb}.

Notably, TSRNet is \textit{trained exclusively on normal }ECG samples but is \textit{tested on both normal and abnormal samples}. Our network follows a reconstruction-by-inpainting anomaly detection approach \cite{zavrtanik2021reconstruction}, where signals with a high restoration error in the inpainting task are identified as anomalies.


\vspace{-1em}
\section{Related Work}
\vspace{-0.5em}
As an important research field, various methods have been devoted to unsupervised time series anomaly detection. DAGMM \cite{dagmm} is an end-to-end method that combines autoencoder with the Gaussian Mixture Model to perform robust anomaly detection in high-dimensional data. MAD-GAN \cite{madgan} focuses on complex multivariate correlations among the time-series data to identify normal and abnormal samples. USAD \cite{usad} follows the adversarial training in encoder-decoder architecture to enhance the reconstruction error of inputs with anomalies while maintaining stability compared to other GAN-based methods. TranAD \cite{tranad} proposes the combination of Transformer and adversarial training procedures to amplify reconstruction errors. Anomaly Transformer \cite{anotran} deals with the prior-association and series-association simultaneously by Anomaly-Attention and minimax optimization strategy. Zheng et al. \cite{tsl} proposes the two-branch CNN with a larger kernel to train the feature extractor, which can create a better anomaly detector. BeatGAN \cite{beatgan} employs time series warping and adversarial regularization to robustly reconstruct the input data.
Unlike the existing works, our work explores the potential of the time-frequency domain, specifically the spectrogram, in unsupervised ECG anomaly detection. This innovative approach harnesses the advantages of both time series and time-frequency representations, mitigating the limitations of relying on a single modality and enabling the extraction of comprehensive ECG signal characteristics for enhanced anomaly detection.

\vspace{-1mm}
\section{Methodology}

In this section, we introduce our unsupervised method, which utilizes multi-lead ECGs (i.e. 12-lead) and their corresponding time-frequency spectrogram. Specifically, our method entails the design of a CNN-based model with cross-attention enhancements, exclusively trained on multimodal normal ECGs through a reconstruction task. Furthermore, we introduce a novel inference method that enhances the accuracy of distinguishing abnormal ECGs from normal ones in the inference phase, termed Peak-based Error.  Our model is designed in an unsupervised manner, and we will provide detailed training and inference process as follows:

\subsection{Training Process}

\begin{figure*}
    \centering
    \includegraphics[width=\linewidth]{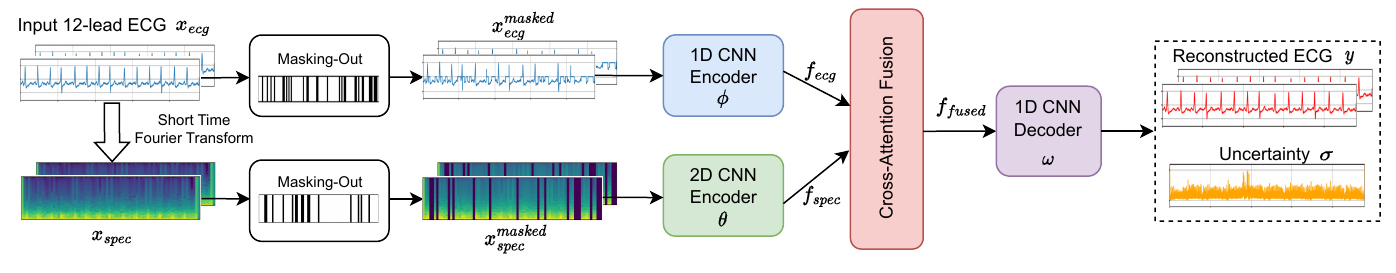}
    \vspace{-2em}
    \caption{The overall of our proposed TSRNet framework: Starting with a 12-lead ECG time-series signal $x_{ecg}$, we generate a time-frequency spectral signal $x_{spec}$ using Short Time Fourier Transform (STFT). Both $x_{ecg}$ and $x_{spec}$ undergo a Masking-Out procedure before being processed by a 1D-CNN encoder $\phi$ and a 2D-CNN encoder $\theta$, resulting in $f_{ecg}$ and $f_{spec}$, respectively. A shared-weight two-layer cross-attention module is proposed to fuse these features, creating a fused feature $f_{fused}$. Finally, a 1D-CNN decoder $\omega$ processes the fused feature $f_{fused}$ to generate the reconstructed ECG signal $y$ and the uncertainty $\sigma$.}
    \label{fig:overall-arch}
\end{figure*}

\noindent
\textbf{Overall Architecture.}
Given a multi-lead ECG signal $x_{ecg}\in\mathbb{R}^{D\times N}$, where $D$ is the signal length and $N$ is the number of leads ($N = 12$ in our case). We employ Short Time Fourier Transform (STFT) \cite{griffin1984signal} on $x_{ecg}$, resulting in the time-frequency spectral signal $x_{spec}\in\mathbb{R}^{D\times N\times H\times W}$. To introduce randomness as well as learn important signal patterns, both $x_{ecg}$ and $x_{spec}$ undergo a Masking-Out procedure (Section \ref{sec:maskout}) to be randomly masked-out. Subsequently, they are separately fed into a 1D-CNN and a 2D-CNN encoders to extract meaningful features $f_{ecg}$ and $f_{spec}$, respectively. 
Next, a shared weight two-layer cross-attention module is tailored to fuse the features $f_{ecg}$ and $f_{spec}$ from the two different domains to form a fused feature $f_{fused}$. Finally, the fused feature $f_{fused}$ is processed by a 1D-CNN decoder to reconstruct the \textit{original ECG signal} and predict the \textit{uncertainty}, which implies the difficulty of reconstructing each data point.
The overall architecture of the model is depicted in Figure \ref{fig:overall-arch}.

\noindent
\textbf{Masking-Out}
\label{sec:maskout}
In this work, we apply a simple yet efficient unsupervised learning approach called inpainting \cite{huang2022self}. This approach involves partially masking the input data to encourage the model to learn crucial signal patterns from the training dataset. For 1D time series data, we apply masking across the sequence by uniformly randomly choosing 30\% data points of each lead. For the 2D spectrogram, instead of masking pixels or square patches, we use the striping approach. That means, we only choose 20\% time intervals, and masking out all values belonging to those intervals. It is noteworthy that the same mask is applied to all the leads. The masked times series and spectrogram are denoted by $x_{ecg}^{masked}$ and $x_{spec}^{masked}$, respectively.

\noindent
\textbf{CNN-based Encoder}
We employ a 1D CNN encoder $\phi$ to distill features from the masked time series data $x_{ecg}^{masked}$ and a 2D CNN encoder $\theta$ to extract frequency-related information from the masked spectrogram $x_{spec}^{masked}$.
\begin{equation}
\begin{split}
    f_{ecg} = \phi(x_{ecg}^{masked})
    \\
    f_{spec}=\theta(x_{spec}^{masked})
\end{split}
\end{equation}
In particular, different leads are treated as different channels of the input.
We do not apply padding in the 2D CNN encoder since padding introduces too much noisy information, making the model perform poorly, whereas padding is still used in the 1D CNN encoder. Each encoder consists of 5 convolutional blocks, including one convolution layer, leaky ReLU activation function, and batch normalization layer.
At the end of $\theta$, we add a 1D convolution layer to transform the 2D latent feature maps into 1D latent feature maps in such a way that the time dimension is preserved. Therefore, the latent features from both the encoders can be treated as sequences of features.

\begin{figure}
    \centering
    \includegraphics[width=\linewidth]{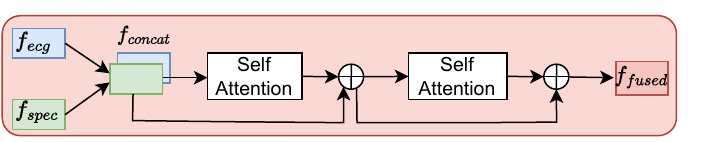}
    \vspace{-2em}
    \caption{Illustration of Cross-Attention Fusion. Two features, $f_{ecg}$ and $f_{spec}$, are concatenated to form $f_{concat}$. Subsequently, this concatenated feature undergoes processing by two shared-weight self-attention layers. $\bigoplus$ denotes the element-wise addition.}
    \label{fig:attention-module}
\end{figure}

\noindent
\textbf{Cross-Attention Fusion}
To combine information from time series and spectrum domains extracted by the encoders $\phi$ and $\theta$, we introduce an innovative cross-attention fusion module to learn robust and informative features as shown in Figure \ref{fig:attention-module}. First, we concatenate the two features $f_{ecg}$ and $f_{spec}$, resulting in feature $f_{concat}$. 
Then we apply two shared weight self-attention layers along with skip connections. The motivation to duplicate the attention layers is to enhance the capacity of the model but not too much so that the model is able to learn the necessary long-range relationship but not too strong to attain generalization ability.
Finally, the processed sequence is channel-wise refined into a shorter sequence by a linear layer $\mathcal{F}$. This can be considered as a regularization technique, preventing the model from reconstructing even noises.
\begin{equation}
\begin{split}
    f_{concat}=concat(f_{ecg}, f_{spec})\\
    f_{att}^1=f_{concat} + att(f_{concat})\\
    f_{att}^2=f_{att}^1 + att(f_{att}^1)\\
    f_{fused}=\mathcal{F}(f_{att}^2)
\end{split}
\end{equation}
In this context, $\mathcal{F}$ comprises normalization and a ReLU activation function. The resulting sequence $f_{fused}$ possesses the same length as that of $f_{ecg}$.

\noindent
\textbf{CNN-based Decoder and Objective Function}
In the final stage, the 1D CNN decoder $\omega$ takes $f_{fused}$ as the input and returns the reconstructed ECG signal $y$, accompanied by an uncertainty $\sigma$ as a measurement of difficulty of reconstructing each data point.

\begin{equation}
    (y, \sigma) = \omega(f_{fused})
\end{equation}
In the training procedure, we optimize the model with the uncertainty-aware restoration loss \cite{ecg-miccai} as follows:
\begin{equation}
\begin{split}
    \mathcal{L}=\frac{\sum_{i=1}^{D}exp(-\sigma^i)\times(y^i-x_{ecg}^i)^2 + \sigma^i}{D}
\end{split}
\end{equation}
Here, $D$ is the number of data points in time-series. Intuitively, we can consider the first term $exp(-\sigma^i)\times(y^i-x_{ecg}^i)^2$ as a constraint on the relationship between the uncertainty $\sigma$ and the squared error $(y-x_{ecg})^2$: if the error is big, then the uncertainty must be also big to mitigate the error. In addition, optimizing the second term $\sigma$ encourages the model to reconstruct the original signal more accurately.

\begin{figure}
    \centering
    \includegraphics[width=\linewidth] {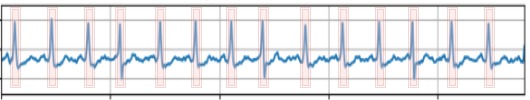}
    \vspace{-2em}
    \caption{Visualization of Peak-based Error in the ECG signal. We only consider the R-peaks when detecting the anomaly signal, which is marked by the pink segments.}
    \label{fig:inference}
\end{figure}

\subsection{Inference with Peak-based Error}
In our training phase, the model exclusively learns from normal ECG signals. Consequently, during the inference phase, any signals associated with anomalies (i.e., unknown classes) will exhibit higher reconstruction errors compared to those from normal signals. As a result, we can directly use the aforementioned uncertainty-aware restoration loss as an indicator of signal anomalies. In other words, the larger the restoration loss $\mathcal{L}$ is, the more likely the signal has an anomaly. However, recognizing that anomalies are often detected around the R-peaks, we introduce a Peak-based Error, defined as follows:

\begin{equation}
    \mathcal{E}=\frac{\sum_{i\in mask}exp(-\sigma^i)\times(y^i-x_{ecg}^i)^2 + \sigma^i}{||mask||}
\end{equation}

Here, the variable $mask$ is defined as a window around R-peaks as shown in  Figure \ref{fig:inference}.
The efficacy of this Peak-based Error strategy is proved by the ablation study shown in the Section \ref{Sec:Experiments}. The classification of an input ECG signal as normal or abnormal is contingent upon $\mathcal{E}$.

\section{Experiments}
\label{Sec:Experiments}
\textbf{A. Dataset and Evaluation Metrics.}

\noindent
\underline{\textit{Dataset:}} We conduct the experiments on PTB-XL dataset \cite{ptb}, including 12-lead ECGs that are 10s in length for each patient. For a fair comparison, we follow the data splitting from the previous work of Jiang et al. \cite{ecg-miccai}, 
which utilizes 8,167 normal ECGs for training, and 912 normal with 1,248 abnormal ECGs for testing.

\noindent 
\underline{\textit{Metrics:}} We evaluate the network’s accuracy using the area under the Receiver Operating Characteristic curve (AUC), while the network’s complexity is measured by the number of trainable parameters (params), and inference time (second).

\noindent
\textbf{B. Implementation Details.} 

\noindent Our model is implemented by the PyTorch library and trained on a single NVIDIA RTX A6000 with 48GB memory. Instead of exhaustively finding an overfitting training procedure, we train our model for 50 epochs, the batch size of 32, the AdamW optimizer \cite{adamw} with an initial learning rate of 1e-4, a weight decay of 1e-5 and cosine learning rate decay scheduler. 


\noindent
\textbf{C. Performance Comparison.} 

\noindent We compare our TSRNet with several SOTA methods in time series anomaly detection, e.g. DAGMM \cite{dagmm}, MAD-GAN \cite{madgan}, USAD \cite{usad}, TranAD \cite{tranad}, Anomaly Transformer \cite{anotran}, Zheng et al. \cite{tsl}, BeatGAN \cite{beatgan}, and Jiang et al. \cite{ecg-miccai}. All the performance results are excerpted from Jiang et al. \cite{ecg-miccai}. Those SOTA methods belong to the unsupervised time series anomaly detection approach with the same setting as our method. As a result, we regard them as the most suitable options for comparison in order to highlight the effectiveness of our proposed method.

Analyzing the results presented in Table \ref{tab:ptb-xl}, it is evident that our TSRNet outperforms other SOTA methods while keeping its model size compact. It is worth highlighting that our model achieves the same peak performance as Jiang et al.'s \cite{ecg-miccai}. However, our model is not only designed as a simpler framework but also requires fewer learned parameters. Consequently, our model boasts significantly faster inference speeds (more than $\times 6$), as demonstrated in Table \ref{tb:compare2}. It’s worth mentioning that the inference time results are averaged over three runs for each method.


\begin{table}[!ht]
\centering
\caption{ECG anomaly detection comparison on PTB-XL dataset. The highest and second highest scores are shown in \textbf{bold} and \underline{underline}, respectively.} \label{tab:ptb-xl}
\resizebox{0.75\linewidth}{!}{
\begin{tabular}{l|c}
\toprule
\textbf{Methods} & \textbf{AUC} $\uparrow$\\
\midrule
    DAGMM (ICLR'18) \cite{dagmm} & 0.782\\
    MAD-GAN (ICANN'19) \cite{madgan} & 0.775\\
    USAD (SIGKDD'20) \cite{usad} & 0.785\\
    TranAD (VLDB'22) \cite{tranad} & 0.788\\
    Anomaly Transformer (ICLR'22) \cite{anotran} & 0.762\\
    Zheng et al. (MICCAI'22) \cite{tsl} & 0.757 \\
    BeatGAN (TKDE'22) \cite{beatgan} & \underline{0.799}\\
    Jiang et al. (MICCAI'23) \cite{ecg-miccai} & \textbf{0.860}\\
\hline
    \textbf{TSRNet (Ours)} & \textbf{0.860}\\
\bottomrule
\end{tabular}}
\vspace{-1.5mm}
\end{table}

\begin{table}[!h]
\centering
\vspace{-1.5mm}
\caption{Detailed comparison between two closely-matched performance methods, i.e. our TSRNet and Jiang et al. \cite{ecg-miccai} on number of trainable parameters (M) and inference time (s).}
\resizebox{0.9\linewidth}{!}{
\begin{tabular}{l|ccc}
\toprule
    \textbf{Methods} & \textbf{AUC} $\uparrow$ & \textbf{Params(M)} $\downarrow$ & \textbf{\shortstack{Inference\\Time(s)}} $\downarrow$\\
\midrule
    Jiang et al. \cite{ecg-miccai} &  \textbf{0.860} & {7.09} &  {0.19} \\
\hline
    \textbf{TSRNet (Ours)} & \textbf{0.860} & \textbf{4.39} & \textbf{0.03} \\
\bottomrule
\end{tabular}}
\label{tb:compare2}
\vspace{-2mm}
\end{table}

\begin{table}[!h]
\vspace{-1mm}
\caption{Ablation study on the effectiveness of each modality (left) and the effectiveness of the proposed Peak-based Error (right).} \label{tab:ablation}
\resizebox{0.5\linewidth}{!}{
\begin{tabular}{ll}
\toprule
\textbf{Modalities}    & \textbf{AUC} $\uparrow$ \\ \midrule
Time-series    &  0.801   \\
Spectrum       &  0.723   \\
\textbf{Combine (Ours) }&   \textbf{0.860 } \\ \bottomrule
\end{tabular}}
\quad
\resizebox{0.45\linewidth}{!}{
\begin{tabular}{cl}
\toprule
\centering
\textbf{Peak-based Error} & \textbf{AUC} $\uparrow$\\ \midrule
\xmark        &  0.852  \\
\cmark       &    \textbf{0.860} \\ \bottomrule
\end{tabular}}
\vspace{-3mm}
\end{table}

\noindent
\textbf{D. Ablation Study.} 

\noindent 
In this ablation study, we assess the effectiveness of our key components. Table \ref{tab:ablation} (left) presents the performance of each individual modality, i.e., time series, spectrogram, and the fusion of both modalities. The results clearly demonstrate that the time series component plays a pivotal role in detecting anomalous ECG patterns, while also acknowledging the positive impact of the spectrogram. Consequently, the combination of both time series and spectrogram data yields the best overall performance. 

Moving to Table \ref{tab:ablation} (right), we assess the efficacy of our proposed Peak-based Error during the inference process. We compare the result when inferring with Peak-based Error to the result when inferring the original objective function \cite{ecg-miccai}. The findings indicate that the inclusion of the Peak-based Error is instrumental in achieving high performance.









\vspace{-1.5mm}
\section{Conclusion}
\vspace{-0.5mm}
\noindent In this study, we introduced TSRNet, a simple and real-time framework designed for ECG anomaly detection. TSRNet is rooted in a restoration-based approach and trained in an unsupervised manner using inpainting mechanism. Notably, TSRNet introduces an innovative perspective by emphasizing the substantial impact of both time-frequency and time-series domains on ECG anomaly detection. Additionally, we introduced an effective cross-attention mechanism to merge information from both modalities, enabling the model to leverage the valuable characteristics embedded in each. For efficient inference, we proposed a Peak-based Error strategy, which prioritizes R-peaks to classify an input ECG signal as either normal or abnormal. The experimental results demonstrate that TSRNet achieves SOTA performance (AUC = 0.860) while maintaining real-time inference capabilities (33.3 fps) and a compact model size (4.39M params).



\section{Acknowledgement}

Nhat-Tan Bui, Thinh Phan, and Ngan Le are supported by the National Science Foundation (NSF) under Award No OIA-1946391 RII Track-1, NSF 1920920 RII Track 2 FEC, NSF 2223793 EFRI BRAID, NSF 2119691 AI SUSTEIN, NSF 2236302.
Minh-Triet Tran is sponsored by Vietnam National University Ho Chi Minh City (VNU-HCM) under grant number DS2020-42-01.
Dinh-Hieu Hoang is funded by Vingroup Joint Stock Company and supported by the Domestic Master/ PhD Scholarship Programme of Vingroup Innovation Foundation (VINIF), Vingroup Big Data Institute (VINBIGDATA), code VINIF.2022.ThS.JVN.04.

\small
\bibliographystyle{IEEEbib}
\bibliography{strings}

\end{document}